\documentclass[apj]{emulateapj}
\usepackage{multirow}

\def\gtorder{\mathrel{\raise.3ex\hbox{$>$}\mkern-14mu
             \lower0.6ex\hbox{$\sim$}}}
\def\ltorder{\mathrel{\raise.3ex\hbox{$<$}\mkern-14mu
             \lower0.6ex\hbox{$\sim$}}}

\slugcomment{Draft of \today}

\begin{document}

\title{Evidence of a Hadronic Origin for the TeV Source J1834$-$087}

\author{D.~A.~Frail\altaffilmark{1}, M.~J.~Claussen\altaffilmark{1},
  \&\ J.~M\'ehault\altaffilmark{2}}

 \altaffiltext{1}{National Radio Astronomy Observatory, P.O. Box O,
   Socorro, NM 87801, USA}

 \altaffiltext{2}{Universit\'e Bordeaux 1, CNRS/IN2P3, Centre d'Etudes
   Nucl\'eaires de Bordeaux Gradignan, 33175 Gradignan, France}


%
\begin{abstract}
%
  We report on the discovery of compact, narrow OH line emission from
  the hydroxl molecule at 1720 MHz toward the extended TeV source
  J1834$-$087. The origin of this high energy emission unknown; it
  could be powered by one or more candidate neutron stars (leptonic)
  or by cosmic rays interacting with dense gas (hadronic). The OH
  emission is detected near the center of J1834$-$087, coincident with
  the radio continuum of the supernova remnant W41, and the radial
  velocity of the line is the same velocity as a giant molecular cloud
  along the line of sight. We argue that the OH is maser emission
  stimulated by the interaction of the W41 shock with the molecular
  cloud. The known correlation between $\gamma$-ray bright supernova
  remnants and OH masers favors a hadronic interpretation for this
  high energy emission.

\end{abstract}


\section{Introduction}\label{sec:Introduction}

The pace of discovery in very high energy (VHE; $>$100 GeV) astronomy
has quickened in recent years thanks to a new generation of
atmospheric Cerenkov imaging telescopes (e.g. H.E.S.S., MAGIC, VERITAS).
These instruments are characterized by a large field-of-view
($\sim5^\circ$), sensitivities of 1\% of the Crab nebula and angular
resolutions of a few arcminutes. Surveys with these instruments have
led to the discovery of nearly 100 VHE galactic and extragalactic
sources; an order of magnitude increase in VHE sources in only a
decade \citep{weekes08,roa13}.

VHE astronomy finds itself in a golden period, not unlike radio
astronomy after World War II, with large catalogs of unidentified
sources. Much effort, therefore, has been going into making
correlated, multi-wavelength observations in order to identify the
sources of VHE emission. Many astrophysical processes are known to
accelerate particles which produce accompanying high energy radiation
and the VHE emission often provides the most stringent tests of
theoretical models.  The sources which have been identified to date
have fallen into several different populations including pulsar wind
nebulae (PWN), shell-type supernova remnants (SNR), binary stars,
Wolf-Rayet stars, giant molecular clouds (GMC), the Galactic Center, nearby
star-forming galaxies, and active galactic nuclei.

Despite this progress, more than 25\% of all VHE sources remain
unidentified. One of the more intriguing of these sources is
J1834$-$087. Discovered by the H.E.S.S. telescope \citep{aab+06}, and
confirmed by the MAGIC telescope \citep{aaa+06}, J1834$-$087 is an
extended TeV source 11$^\prime$ diameter (see Fig. 1). There is a
likely GeV counterpart 2FGL J1834.3$-$0848 detected by the {\it Fermi}
satellite \citep{naa+11}. The extended VHE emission is localized to
near the center of the large diameter (27$^\prime$) shell-type SNR W41
or G23.3$-$0.3 \citep{hbw+06}. These basic properties are confirmed
with a deeper H.E.S.S. integration (52 hrs) and a re-analysis of 4
years of Fermi data \citep{mtr+11}. The distance to W41 has been
determined to be approximately 4.2 kpc \citep{lt08}.

There is no consensus on the origin of the high energy emission from
J1834$-$087. The gamma-rays could be produced from a leptonic process,
whereby the relativistic electrons accelerated at the termination
shock of the pulsar wind nebulae up-scatter low energy photons via
inverse Compton scattering. There are no fewer than three candidate
neutron stars that might be responsible. \citet{bb08} suggest that the
Vela-like PSR J1833$-$0827, 24$^\prime$ from W41, is responsible for
the VHE emission. XMM observations toward J1834$-$087 detect a bright,
hard X-ray source and faint, diffuse emission offset from the point
source \citep{mgh09}. The authors argue that the X-ray emission
originates from a young pulsar and its accompanying PWN. Follow-up
observations made with the {\it Chandra} satellite found no evidence
for a physical association between the point source and the extended
emission \citep{mkp11}. Instead, the compact XMM source is resolved
into a point source CXOU J183434.9$-$084443 surrounded by diffuse
emission 20$^{\prime\prime}$ in diameter, whose spectrum is more
suggestive of dust scattering than a pulsar wind nebula. No X-ray
pulsations have been detected from the point source.  Finally, there
is the magnetar Swift J1834.9-0846, seen in projection towards
J1834$-$087 \citep{kkp+12}, but no claims have been made that this
magnetized neutron star could be powering the extended TeV emission.

Alternatively, the TeV emission from J1834$-$087 may have a hadronic
origin. Cosmic rays, accelerated by a supernova remnant collide with
the ambient gas producing neutral pions which decay into gamma-rays
($\pi^\circ\rightarrow\gamma\gamma$). Strong evidence in favor of the
hadronic origin has recently been found in a number of galactic SNRs
\citep{gct+11, aaa+13}. \citet{aaa+06} suggested a hadronic origin for
J1834$-$087, noting the spatial coincidence of a GMC toward the center
of W41. A similar morphological argument was put forth by
\citet{tllw07}, based on CO and HI spectra. Neither of these studies,
however, provide any {\it direct} evidence that the SNR shock of W41
is interacting with a molecular cloud. In this paper we undertake
observations of the hydroxl molecule (OH) in its ground state at
1720.530 MHz (\S{\ref{SurveyObs}}).  Through repeated studies, OH(1720
MHz) masers have emerged as an important tracer of molecular shocks,
and their detection (\S\ref{Discussion}) can be taken to be strong
evidence in favor of gamma-rays from hadronic particle acceleration.

\section{Observations}\label{SurveyObs}

The Karl G. Jansky Very Large Array (VLA) observed toward J1834$-$087
on 2011 October 25 under program 11B-009. Three pointings were made,
each with a 30-arcmin diameter field of view. The first pointing was
at the center of W41 ($l,b=23.3^\circ,-0.3^\circ$). The second
pointing was toward the supernova remnant G22.7$-$0.2
($l,b=22.5^\circ,-0.1^\circ$), which is adjacent to W41 and may be
co-located. The third pointing was made between the two SNRs
($l,b=23.0^\circ,-0.2^\circ$). The VLA was used in its D
configuration, giving a synthesized beam of approximately 45-arcsec.

The VLA WIDAR correlator was configured to collect data in dual
polarization, with 256 channels each of width 3.906 kHz (0.7 km
s$^{-1}$). We observed the ground state OH line at a rest frequency of
1720.530 MHz, shifting the center of the observing band to an LSR
velocity of 70 km s$^{-1}$. This experiment is sensitive to OH lines with
velocities from $-15$ km s$^{-1}$ to +155 km s$^{-1}$.

The data was calibrated in the Astronomical Image Processing System
(AIPS) package following standard practice. The radio continuum was
subtracted in the visibility plane before making spectral line data
cubes. Neither deconvolution nor self-calibration were performed on
these images. The rms noise in the final three images varied between
10 and 11 mJy beam$^{-1}$.

\begin{figure}
\centerline{\includegraphics[angle=0,scale=0.46]{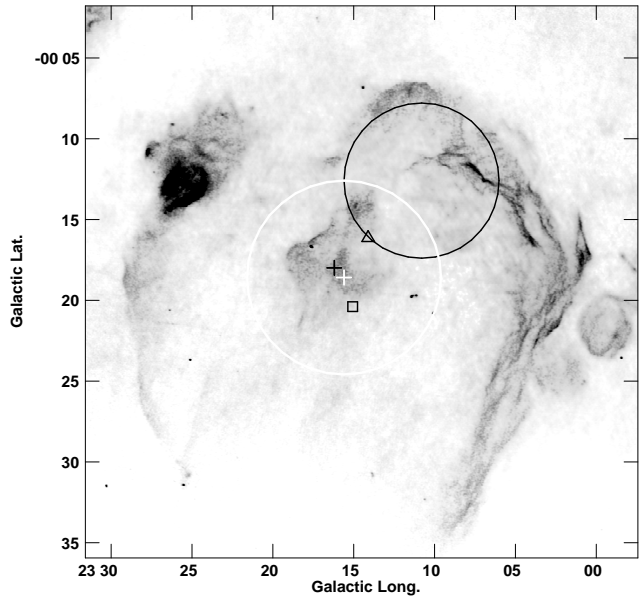}}
\caption{A 20-cm radio image (greyscale) of the SNR W41 taken from the
  Multi-Array Galactic Plane Imaging Survey (MAGPIS). This image is
  displayed in Galactic coordinates and is centered on the extended
  VHE source, whose position (l,b=23.26$^\circ$,$-0.31^\circ$) and
  size (radius=0.20$^\circ$) are indicated by the large circle. The
  smaller circle (radius=0.16$^\circ$) indicates the centroid of the
  extended Fermi emission at (l,b=23.18$^\circ$,$-0.21^\circ$). The
  square and triangle mark the positions of Swift J1834.9-0846 and CXOU
  J183434.9$-$084443, respectively, while the two crosses mark the
  locations of the OH maser candidates.  See text for more details.}
\label{Fig:one}
\end{figure}

\section{Results}\label{results}

We searched for lines in the Stokes I image cubes using the Common
Astronomy Software Applications (CASA) package. No OH lines were
detected toward the second or third pointings, i.e. toward the SNR
G22.7$-$0.2 or at the interface between W41 and G22.7$-$0.2. However,
there is at least one compact OH emission feature in the W41 pointing
at R.A.(J2000)=18h 34m 47.44s, Dec.(J2000)=$-$8$^\circ$ 44$^\prime$
18.7$^{\prime\prime}$ ($\pm 3.5^{\prime\prime}$), or equivalently in
Galactic coordinates $l,b$=(23.26$^\circ$, $-$0.31$^\circ$), with a peak
flux density of S$_p$=64 mJy, a velocity V$_{\rm LSR}$=74.2 km
s$^{-1}$ and a line width $\Delta$V=1.3 km s$^{-1}$. There is a second
weaker feature with S$_p$=59 mJy, a velocity V$_{\rm LSR}$=74.9 km
s$^{-1}$ and a line width $\Delta$V=2.6 km s$^{-1}$. At
R.A.(J2000)=18h 34m 46.00s, Dec.(J2000)=$-$8$^\circ$ 43$^\prime$
58.6$^{\prime\prime}$ ($\pm 3.5^{\prime\prime}$) or
$l,b$=(23.27$^\circ$, $-$0.30$^\circ$), this feature is only
29$^{\prime\prime}$ away from the brighter line. Formally these fit as
two separate lines but the synthesized beam is rather large
51$^{\prime\prime}\times 37^{\prime\prime}$ (15$^\circ$), and higher
angular resolution is needed to clearly separate them. In addition to
these emission spots, we detect diffuse emission and absorption. Two
arcminutes due north of the brightest line there is a
90$^{\prime\prime}$ long ridge of faint emission at V$_{\rm LSR}$=71
km s$^{-1}$ and corresponding absorption at V$_{\rm LSR}$=76 km
s$^{-1}$. The velocity width of the absorption features is
approximately twice that of the emission. There is no OH emission or
absorption detected elsewhere in the 30-arcmin field centered on W41.


Diffuse OH emission (and absorption) at 1720 MHz has been detected
toward a number of SNRs \citep{yur95,ygr+99}, while compact,
narrow-line OH emission has been seen now toward 24 galactic SNRs
\citep{frail11}.  These compact features have been identified as
simulated emission from a maser process based on their narrow lines
and high brightness temperatures. Establishing a maser origin is
important, for as we show in \S\ref{Discussion} the excitation of
OH(1720 MHz) masers {\it requires} a collisional pump not a radiative
pump. This distinguishes OH(1720 MHz) masers from the other satellite
line at 1612 MHz and the mainline OH masers at 1665 and 1667 MHz,
which all can be radiatively pumped. Unfortunately, we cannot
unambiguously identify the compact OH line toward W41 as maser
emission. While the narrow linewidth is indicative of OH masers, the
low angular resolution of our VLA observations only allows us to put a
lower limit on the brightness temperature of T$_b\gtorder$20 K.
Existing observations have shown the OH (1720 MHz) masers to be
$\sim$200 AU in size with T$_b\simeq 10^6$ K, with a central core of
T$_b \simeq 10^8$ K \citep{hgb+03}. Some of the diffuse OH emission
may also be due to masers.  \citet{hyw08} showed that there exists
widespread, low-gain maser emission T$_b\ltorder$2500 K toward these
SNRs that had been resolved out by interferometer surveys.  Future
high angular resolution observations can easily confirm whether the OH
emission toward W41 is due to masers or not.

If the compact, narrow-line features are masers, then they lie at the
faint end of the flux density distribution of known OH(1720 MHz)
masers where \citet{hy09} previously noted that existing surveys are
incomplete. The W41 maser luminosity, defined as S$_p\times{d}^2$
(where $S_p$ is the peak flux density in mJy and $d$ is the distance
in kpc), is 10$^3$ mJy kpc$^2$. This value is fainter than the median
of 10$^4$ mJy kpc$^2$ but this distribution spans some four orders of
magnitude \citep{gfgo97}.

The putative masers are detected near the center of J1834$-$087,
coincident with the radio continuum of W41, which is interior to the
outer shell of the SNR (see Fig. 1). The maser positions are also
coincident within the errors ($\pm0.01^\circ$) of the centroid of the
newly revised extended H.E.S.S. position of $l,b$=(23.26$^\circ$,
$-$0.31$^\circ$), implying a direct relation between the VHE emission
and the masers \citep{mtr+11}. The radial velocity of the OH emission
(V$_{\rm LSR}$=74 km s$^{-1}$) coincides with velocity centroid of a
GMC noted by \citet{aaa+06} and \citet{lt08}. This velocity agreement
establishes that that the OH emission originates from the GMC.  There
is some additional kinematic support for the SNR interacting with the
GMC. The radio continuum of W41 is seen in absorption against OH has
at V$_{\rm LSR}$=76 km s$^{-1}$, and thus the SNR lies behind the GMC
at least.  Later in \S\ref{Discussion} we show more directly that the
detection of OH requires that the SNR is interacting with the
molecular cloud.

\section{Discussion}\label{Discussion}

We now present the argument that the detection of OH emission at 1720
MHz toward J1834$-$087 favors a hadronic origin for the VHE emission.
Specifically, the detection of this line is evidence that the SNR W41
is interacting with the adjacent molecular cloud and that the
association is not a line of sight coincidence. Ever since the
original detection of OH(1720 MHz) masers in the SNR W28
\citep{fgs94}, this satellite transition of the hydroxl molecule has
become a powerful tool for identifying when a SNR shock is interacting
with a molecular cloud \citep[see reviews by][]{wy02,frail11}.
OH(1720 MHz) masers are found in about 10\% of all Galactic SNRs and
they are located on or near the synchrotron peaks. Observations of
other molecular species have shown that the OH(1720 MHz) masers are
found coincident with dense gas with broad line widths, indicative of
post-shock gas \citep{hrar09, rrj05, atst99, fm98, km97}. The velocity
of the masers matches the systemic velocity of the molecular gas
because strong maser amplification favors shocks transverse shocks
with long coherent pathlengths \citep{gfgo97}.

It is thought that the OH molecule is formed downstream of a
non-dissociative, compression-type shock (20-30 km s$^{-1}$) that has
propagated into a molecular cloud with densities of order
$n_\circ=$10$^4$-10$^{5}$ cm$^{-3}$ and temperatures of 30-120 K.
Under these conditions a strong collisionally pumped maser transition
results for the OH molecule at 1720 MHz \citep{lge99}. Thus the
detection of a OH(1720 MHz) maser is unambiguous proof of an
interaction, unlike other tracers such as HI or CO emission. The OH
provides other physically useful information like the gas density, the
geometry of the shock, the radial velocity (for kinematic distances),
and in some cases the {\it in situ} magnetic field.

\cite{cfgg97} first noted the link between $\gamma$-ray bright SNRs
\citep{ehk+96} and OH(1720 MHz) masers. This correlation has only
gotten stronger as the number of TeV/GeV SNRs has grown
\citep{hyw08,cs10}.  Prominent SNRs with OH(1720 MHz) masers and
bright $\gamma$-rays include W28, W44, IC443, W51C and W30 \citep[see
tables in][]{frail11,hvb+12}.  The physical relationship between
OH(1720 MHz) masers and $\gamma$-ray bright SNRs may be direct or
indirect. In the later case the OH is produced downstream of the
molecular shock by dissociating H$_2$O using the thermal X-rays
interior to the remnant \citep{war99} and hence the maser is just a
signpost for the SNR interacting with a molecular cloud. In the later,
it is the enhanced production of the cosmic rays at the SNR shock that
dissociates the H$_2$O into OH \citep{hyw08}. In this case there is a
close one-to-one relationship between OH and cosmic rays. Observations
of molecular ions serve as a diagnostic of cosmic ray ionization rates
$\zeta_{\rm CR}$ and therefore could test this hypothesis directly.
Recent measurements of $\zeta_{\rm CR}$ near SNRs show enhanced values
in dense molecular clouds adjacent to the W51C \citep{chm+11} and in
diffuse clouds adjacent to IC443 \citep{ibg+10}.

For the specific example of J1834$-$087, the coincidence of TeV
emission toward a GMC does not require that the W41 SNR be the
particle accelerator \citep{aaa+06}. Likewise, there is no strong
evidence that W41 is interacting with the ubiquitous neutral (HI) or
molecular (CO) gas along the line of sight \citep{tllw07}. However,
with the detection of the OH at the same velocity of the molecular
cloud, there is now stronger evidence that support these earlier
claims that the SNR is interacting with the GMC. The shock-excited
OH(1720 MHz) maser emission requires that the SNR interacts with the
same dense molecular gas needed to produce bright $\gamma$-ray
emission.  This interaction reproduces the observed gamma-ray
luminosity provided the W41 SNR was the result of a canonical
10$^{51}$ erg explosion giving up a few percent of its energy to
cosmic ray acceleration \citep{tllw07}. W41 joints a group of
middle-aged, $\gamma$-ray bright SNRs including W44, IC443, W28, and
W51C. Their interaction with a clumpy, dense molecular cloud is key to
explaining their gamma-ray luminosity \citep{chev99,ubftt10} and other
unusual properties, including steep VHE spectra \citep{cap11}.

\citet{mkp11} argue that the location of J1834$-$087, within W41 is a
challenge for the hadronic model. They expect the brightest high
energy emission would originate from the outer shell of W41 where the
synchrotron emission (Fig. 1) traces the interaction with the
surrounding medium. Note also that the GeV emission from 2FGL
J1834.3$-$0848 does partially overlap with the W41 shell.  Either the
offsets between the GeV and TeV emission are real, or they originate
due to uncertainties in correcting for the diffuse background and the
large point spread function of {\it Fermi} compared to H.E.S.S.
Nonetheless, if we compare W41 to the sample of SNRs with show a
correlation $\gamma$-rays and OH(1720 MHz) masers this offset does not
appear to be a significant problem for the model. While OH masers are
seen along SNR shells, they are not uncommon interior to SNRs.
J1834$-$087 and the OH masers are found toward a bright, extended
synchrotron emission and in this respect the W41 system resembles
IC443 in terms of the location of the masers and the GeV/TeV emission
\citep{tgc+10,aaa+09}.

We end by noting that while these observations give new impetus to
hadronic models, they do not diminish the possibility that some of the
VHE emission from J1834$-$087 could be leptonic in origin produced by
one or more energetic neutron star candidates. Further radio
observations could strengthen the hadronic model, improving the limits
on the brightness temperature of the OH emission and perhaps measuring
the {\it in situ} magnetic field using Zeeman splitting. A refinement
in the properties of the dense gas would also be possible by imaging
J1834$-$087 lines such as NH$_3$ and SiO, for example \citep{nrb+12,
  bgh+13}. The detection of OH masers indicates post-shock molecular
gas with densities and temperatures over a broad range of order
$n_\circ=$10$^4$-10$^{5}$ cm$^{-3}$ and 30-120 K. However, an accurate
mapping of dense gas could lead to better quantitative estimates of
expected brightness of the high energy emission produced by W41.

\acknowledgments

The National Radio Astronomy Observatory is a facility of the National
Science Foundation operated under cooperative agreement by Associated
Universities, Inc. This research has made use of NASA's Astrophysics
Data System.


\begin{thebibliography}{}


\bibitem[Ackermann et al.(2013)]{aaa+13} Ackermann, M., et al.\ 2013,
  Science, 339, 807

\bibitem[Acciari et al.(2009)]{aaa+09} Acciari, V.~A., Aliu, 
E., Arlen, T., et al.\ 2009, \apjl, 698, L133 


\bibitem[Ajello et al.(2012)]{aab+12} Ajello, M., Allafort, 
A., Baldini, L., et al.\ 2012, \apj, 744, 80 

\bibitem[Albert et al.(2006)]{aaa+06} Albert, J., Aliu, E., Anderhub,
  H., et al.\ 2006, \apjl, 643, L53


\bibitem[Aharonian et al.(2006)]{aab+06} Aharonian, F., Akhperjanian,
  A.~G., Bazer-Bachi, A.~R., et al.\ 2006, \apj, 636, 777

\bibitem[Arikawa et al.(1999)]{atst99} Arikawa, Y., Tatematsu, K.,
  Sekimoto, Y., \& Takahashi, T.\ 1999, \pasj, 51, L7


\bibitem[Bartko \& Bednarek(2008)]{bb08} Bartko, H., \& Bednarek,
  W.\ 2008, \mnras, 385, 1105

\bibitem[Brogen et al. (2013)]{bgh+13} Brogan et al.\ 2013, ApJ submitted 

\bibitem[Caprioli(2011)]{cap11} Caprioli, D.\ 2011, Journal of
  Cosmology and Astroparticle Physics, 5, 26


\bibitem[Ceccarelli et al.(2011)]{chm+11} Ceccarelli, C., 
Hily-Blant, P., Montmerle, T., et al.\ 2011, \apjl, 740, L4 

\bibitem[Chevalier(1999)]{chev99} Chevalier, R.~A.\ 1999, 
\apj, 511, 798 

\bibitem[Claussen et al.(1997)]{cfgg97} Claussen, M.~J., Frail, D.~A.,
  Goss, W.~M., \& Gaume, R.~A.\ 1997, \apj, 489, 143

\bibitem[Castro \& Slane(2010)]{cs10} Castro, D., \& Slane, P.\ 2010,
  \apj, 717, 372


\bibitem[Esposito et al.(1996)]{ehk+96} Esposito, J.~A., Hunter,
  S.~D., Kanbach, G., \& Sreekumar, P.\ 1996, \apj, 461, 820

\bibitem[Frail(2011)]{frail11} Frail, D.~A.\ 2011, \memsai, 82, 
703 

\bibitem[Frail et al.(1994)]{fgs94} Frail, D.~A., Goss, W.~M., \&
  Slysh, V.~I.\ 1994, \apjl, 424, L111

\bibitem[Frail \& Mitchell(1998)]{fm98} Frail, D.~A., \& Mitchell,
  G.~F.\ 1998, \apj, 508, 690


\bibitem[Giuliani et al.(2011)]{gct+11} Giuliani, A., Cardillo, M.,
  Tavani, M., et al.\ 2011, \apjl, 742, L30

\bibitem[Green et al.(1997)]{gfgo97} Green, A.~J., Frail, 
D.~A., Goss, W.~M., \& Otrupcek, R.\ 1997, \aj, 114, 2058 

\bibitem[Helder et al.(2012)]{hvb+12} Helder, E.~A., Vink, J., 
Bykov, A.~M., et al.\ 2012, \ssr, 173, 369 


\bibitem[Helfand et al.(2006)]{hbw+06} Helfand, D.~J., Becker, R.~H.,
  White, R.~L., Fallon, A., \& Tuttle, S.\ 2006, \aj, 131, 2525

\bibitem[Hewitt et al.(2008)]{hyw08} Hewitt, J.~W., 
Yusef-Zadeh, F., \& Wardle, M.\ 2008, \apj, 683, 189 

\bibitem[Hewitt et al.(2009)]{hrar09} Hewitt, J.~W., Rho, J., 
Andersen, M., \& Reach, W.~T.\ 2009, \apj, 694, 1266 

\bibitem[Hewitt \& Yusef-Zadeh(2009)]{hy09} Hewitt, J.~W., \&
  Yusef-Zadeh, F.\ 2009, \apjl, 694, L16

\bibitem[Hoffman et al.(2003)]{hgb+03} Hoffman, I.~M., Goss, W.~M.,
  Brogan, C.~L., Claussen, M.~J., \& Richards, A.~M.~S.\ 2003, \apj,
  583, 272

\bibitem[Indriolo et al.(2010)]{ibg+10} Indriolo, N., Blake,
G.~A., Goto, M., Usuda, T., Oka, T., Geballe, T.~R., Fields, B.~D.,
\& McCall, B.~J.\ 2010, \apj, 724, 1357


\bibitem[Kargaltsev et al.(2012)]{kkp+12} Kargaltsev, O., Kouveliotou,
  C., Pavlov, G.~G., et al.\ 2012, \apj, 748, 26

\bibitem[Koo \& Moon(1997)]{km97} Koo, B.-C., \& Moon, D.-S.\ 1997,
  \apj, 485, 263


\bibitem[Leahy \& Tian(2008)]{lt08} Leahy, D.~A., \&
  Tian, W.~W.\ 2008, \aj, 135, 167

\bibitem[Lockett et al.(1999)]{lge99} Lockett, P., Gauthier,
E., \& Elitzur, M.\ 1999, \apj, 511, 235

\bibitem[M\'ehault et al.(2011)]{mtr+11} M\'ehault et al., 2011,
International Cosmic Ray Conference (ICRC) Proceedings.

\bibitem[Misanovic et al.(2011)]{mkp11} Misanovic, Z., Kargaltsev, O.,
  \& Pavlov, G.~G.\ 2011, \apj, 735, 33

\bibitem[Mukherjee et al.(2009)]{mgh09} Mukherjee, R., 
Gotthelf, E.~V., \& Halpern, J.~P.\ 2009, \apj, 691, 1707 

\bibitem[Nicholas et al.(2012)]{nrb+12} Nicholas, B.~P., Rowell, G.,
  Burton, M.~G., et al.\ 2012, \mnras, 419, 251

\bibitem[Nolan et al.(2012)]{naa+11} Nolan, P.~L., Abdo, A.~A.,
  Ackermann, M., et al.\ 2012, \apjs, 199, 31

\bibitem[Reach et al.(2005)]{rrj05} Reach, W.~T., Rho, J., \& Jarrett,
  T.~H.\ 2005, \apj, 618, 297

\bibitem[Rieger et al.(2013)]{roa13} Rieger, F.~M., de 
Ona-Wilhelmi, E., \& Aharonian, F.~A.\ 2013, arXiv:1302.5603 

\bibitem[Tavani et al.(2010)]{tgc+10} Tavani, M., Giuliani, 
A., Chen, A.~W., et al.\ 2010, \apjl, 710, L151 


\bibitem[Tian et al.(2007)]{tllw07} Tian, W.~W., Li, Z., 
Leahy, D.~A., \& Wang, Q.~D.\ 2007, \apjl, 657, L25 

\bibitem[Uchiyama et al.(2010)]{ubftt10} Uchiyama, Y., Blandford,
  R.~D., Funk, S., Tajima, H., \& Tanaka, T.\ 2010, \apjl, 723, L122

\bibitem[Wardle(1999)]{war99} Wardle, M.\ 1999, \apjl, 525,
L101

\bibitem[Wardle \& Yusef-Zadeh(2002)]{wy02} Wardle, M.,
  \& Yusef-Zadeh, F.\ 2002, Science, 296, 2350

\bibitem[Weekes(2008)]{weekes08} Weekes, T.~C.\ 2008, American 
Institute of Physics Conference Series, 1085, 3 

\bibitem[Yusef-Zadeh et al.(1995)]{yur95} Yusef-Zadeh, F., Uchida,
  K.~I., \& Roberts, D.\ 1995, Science, 270, 1801

\bibitem[Yusef-Zadeh et al.(1999)]{ygr+99} Yusef-Zadeh, F., 
Goss, W.~M., Roberts, D.~A., Robinson, B., 
\& Frail, D.~A.\ 1999, \apj, 527, 172 


\end{thebibliography}
\end{document}